\documentclass[11pt]{article}
\usepackage{graphicx} 
\usepackage{amsmath}
\usepackage{amsfonts}
\usepackage{amssymb}
\usepackage{caption}
\usepackage{geometry}
\usepackage{authblk}
\title{Comment on ``Emergent Gravity and the Dark Universe''
by Erik Verlinde}
\author{Youngsub Yoon$^1$, Ho Seong Hwang$^{2,3}$}

\affil{\emph{$^1$Department of Physics and Astronomy, Sejong University,} \protect\\ \emph{209 Neungdong-ro Gwangjin-gu, Seoul 05006, Republic of Korea}}
\affil{\emph{$^2$Department of Physics and Astronomy, Seoul National University,}\protect\\\emph{Gwanak-gu, Seoul 08826, Republic of Korea}}
\affil{\emph{$^3$SNU Astronomy Research Center, Seoul National University,}\protect\\\emph{Seoul 08826, Republic of Korea}}

\begin{document}

\maketitle
\begin{abstract}
Verlinde suggested a new theory of gravity called ``emergent gravity,'' which resembles Modified Newtonian Dynamics, the alternative to dark matter theory. For his version of Milgrom's constant, he theoretically derived $a_M=cH_0/6=1.1\times 10^{-10}$m/s$^2$ by assuming that our universe is a flat de Sitter space, which is not certainly true. In 2022, when Park and us applied Verlinde's emergent gravity to galaxy rotation curves, we discovered that a slightly smaller value of $a_M$ is preferred. We re-ran our codes and obtained that a value about 30\% smaller than Verlinde's original value of Milgrom's constant is most preferred. This agrees with the value obtained recently by applying Verlinde's emergent gravity to the general FLRW universe.

\end{abstract}

\section{Introduction}
Verlinde's emergent gravity tries to answer the missing mass problem without supposing dark matter by suggesting that Newtonian gravity fails in low gravitational accelerations \cite{Verlinde}. Like Modified Newtonian Dynamics (MOND), it can explain the Tully-Fisher relation, the key empirical relation in galaxy rotation curves, as Verlinde showed himself. In Modified Newtonian Dynamics, Milgrom's constant $a_M$, given by $1.2\times 10^{-10}$m/s$^2$ plays a central role. It is the acceleration scale, for which Newtonian gravity begins to fail, and it also appears in the Tully-Fisher relation. What is remarkable about Verlinde's feat is that he obtained the value of $a_M$ in terms of the Hubble constant. According to him, it is given by $cH_0/6$, which is about $1.1\times 10^{-10}$m/s$^2$. Thus, he connected the accelerated expansion of our universe with galaxy rotation curves, which have no obvious relation from the point of view of MOND.

In 2022, Park and us tested Verlinde's emergent gravity by applying it to galaxy rotation curves \cite{galaxy}. We found that Verlinde's emergent gravity explains the galaxy rotation curves well, but a somewhat lower value of $a_M$ than $cH_0/6$ is preferred. Nevertheless, we did not check exactly which value fits the galaxy rotation curves the best.

In this study, we ran our codes for the galaxy rotation curves again and found out that a value that is about 30\% smaller than Verlinde's value is most preferred for Milgrom's constant. This agrees with the value obtained recently by applying Verlinde's emergent gravity to the general FLRW universe.

The organization of this paper is as follows. In Section \ref{sectionbrief}, we briefly introduce Verlinde's emergent gravity. In Section \ref{sectioncomparison}, we remark that galaxy rotation curves fit the best for the value of Milgrom's constant smaller than Verlinde's original value by approximately 30\%. In Section \ref{FLRW}, we mention that the same value, a value smaller than Verlinde's value by about  30\%, is also obtained by applying Verlinde's emergent gravity to the general FLRW universe. In Section \ref{sectionconclusion}, we conclude our paper.

\section{Very brief introduction to Verlinde's emergent gravity}\label{sectionbrief}
In 2011, Verlinde suggested ``entropic gravity'' \cite{entropic}. He argued that gravity is an entropic force. Let us explain what it is. A system always evolves into a configuration that has a higher entropy. Therefore, it tends to move toward a position that has higher entropy. Then, such a movement may apparently seem to be due to a force, but it is actually only due to entropy. This is called the entropic force, a well-known concept in polymer physics.

By assuming the Bekenstein-Hawking entropy for the entropy associated with area, Verlinde successfully derived Newton's law of gravitation and the Einstein field equation. In particular, the fact that the entropy is proportional to area results in Newton's inverse square law. However, it is not possible to prove entropic gravity by experiments, as its theoretical prediction is exactly the same as Newtonian gravity and general relativity.

In 2017, Verlinde went further by suggesting emergent gravity \cite{Verlinde}. He assumed that our universe is a flat de Sitter space. De sitter space has an event horizon, thus, also an associated Bekenstein-Hawking entropy. Verlinde assumed that the de Sitter entropy is uniformly distributed in our universe. In other words, if there is a volume, it contains entropy proportional to the volume, due to the uniformly distributed de Sitter entropy. This volume entropy competes with the area entropy introduced in 2011. For a large scale, the volume entropy is relatively important compared to the area entropy, because the volume entropy scales as $r^3$, while the area entropy only as $r^2$. Therefore, the gravitational acceleration deviates from the inverse square law, because the volume entropy is no longer negligible. In particular, if our universe is a flat de Sitter space, the volume entropy is given by 
\begin{equation}
\frac SV = \frac{3H_0}{4G}.    \label{SV}
\end{equation}
Milgrom's constant is proportional to the volume entropy.

\section{Comparison with galaxy rotation curves}\label{sectioncomparison}
In \cite{galaxy}, Park and us fitted galaxy rotation curves with Verlinde's emergent gravity. We assumed $a_M=a_0/6$ and considered two values of $a_0$: de Sitter value, which is $a_0=cH_0=6.7\times 10^{-10}$m/s$^2$, the one obtained by Verlinde by assuming our universe is a flat de Sitter space, and quasi de Sitter value, $a_0=5.4\times 10^{-10}$m/s$^2$, which was considered in \cite{quasideSitter}. Park and us found that the quasi de Sitter value, which is slightly smaller than the de Sitter value, fits galaxy rotation curves better. We repeated our earlier analysis by changing $a_0$. See Table \ref{table}. We see that $a_0=4.55\times 10^{-10}$m/s$^2$ fits the best. In this case, we have $a_0/6=0.76\times 10^{-10}$m/s$^2$, which is smaller than Verlinde's value by about 30\%. 

\begin{table}[ht!]
	\begin{center}
		\begin{tabular}{|l|r|r|r|r|}
			\hline
			$a_0$ ($10^{-10}$m/s$^2$)& $\mu$  & $\mu_{\rm err}$  & $\sigma$  & $\sigma_{\rm err}$ \\ 
			\hline
			4.7 & -0.005 & 0.002 & 0.130 & 0.002   \\
			4.6 & -0.001 & 0.003 & 0.128 & 0.003     \\
			4.55 & 0.000 & 0.003 & 0.128 & 0.003   \\
			4.5 & 0.002 & 0.003 & 0.128 & 0.003  \\
                4.4 &   0.005   &   0.003   & 0.128    &  0.003       \\
			\hline
		\end{tabular}
		\caption{Galaxy rotation curve fit for various $a_0$. $\mu$ and $\sigma$ denotes the mean and the standard deviation of $\log_{10} (g_{\rm obs}/g_{\rm Ver})$ where $g_{\rm obs}$ is the observed gravitational acceleration and $g_{\rm Ver}$ is the Verlinde predicted gravitational acceleration. We see that $a_0=4.55\times 10^{-10}$m/s$^2$ fits the best.}
		\label{table}
	\end{center}
\end{table}

\section{Comparison with the FLRW model for Verlinde's emergent gravity}\label{FLRW}
In \cite{opendeSitter}, one of us discarded Verlinde's original assumption that our universe is a flat de Sitter space, as the observed Hubble parameter certainly does not remain constant over time, and only assumed that our universe is described by the FLRW metric. Then, from the observed age of the universe, the parameters for our universe were obtained, which, in turn, showed great agreement with the observed Hubble parameters at various redshifts. Subsequently, the following volume entropy was calculated by considering the obtained parameters of the universe:
\begin{equation}
    \frac SV\approx 0.5348 \frac{H_0}{G}.
\end{equation}
This value is about 30\% smaller than (\ref{SV}), which implies that the real value of Milgrom's constant must be approximately 30\% smaller than Verlinde's original value $cH_0/6$. This agrees with the result of the last section that Milgrom's constant must be smaller than Verlinde's value by about 30\%.

\section{Conclusion}\label{sectionconclusion}
In this manuscript, we mentioned that Milgrom's constant obtained by applying Verlinde's emergent gravity to galaxy rotation curves agrees with the one obtained theoretically, by fitting the parameters in the theory to the observed age of the universe. They are both smaller than Verlinde's value $cH_0/6$ by about 30\%. This shows that Verlinde's emergent gravity passed a crucial test: The observed anomalies in galaxy rotation curves have the common origin as the expansion of our universe. Finally, we mention that there can be a slight difference between the value of Milgrom's constant in Modified Newtonian Dynamics and the one in Verlinde's emergent gravity, as the key equations of the two theories are more or less different.

\section*{Acknowledgement}
This work was supported by the National Research Foundation of Korea [NRF-2021R1A2C1094577(HSH), NRF-2022R1A2C1092306(YY)]. H.S.H also acknowledges the support of Samsung Electronic Co., Ltd. (Project Number IO220811-01945-01), and Hyunsong Educational \& Cultural Foundation.


\begin{thebibliography}{9}
\bibitem{Verlinde}
E.~P.~Verlinde,
``Emergent Gravity and the Dark Universe,''
SciPost Phys. \textbf{2}, no.3, 016 (2017)
doi:10.21468/SciPostPhys.2.3.016
[arXiv:1611.02269 [hep-th]].

\bibitem{galaxy}
Y.~Yoon, J.~C.~Park and H.~Seong Hwang,
``Understanding galaxy rotation curves with Verlinde\textquoteright{}s emergent gravity,''
Class. Quant. Grav. \textbf{40}, no.2, 02LT01 (2023)
doi:10.1088/1361-6382/acaae6
[arXiv:2206.11685 [gr-qc]].

\bibitem{entropic}
E.~P.~Verlinde,
``On the Origin of Gravity and the Laws of Newton,''
JHEP \textbf{04}, 029 (2011)
doi:10.1007/JHEP04(2011)029
[arXiv:1001.0785 [hep-th]].


\bibitem{quasideSitter}
A.~Diez-Tejedor, A.~X.~Gonzalez-Morales and G.~Niz,
``Verlinde\textquoteright{}s emergent gravity versus MOND and the case of dwarf spheroidals,''
Mon. Not. Roy. Astron. Soc. \textbf{477}, no.1, 1285-1295 (2018)
doi:10.1093/mnras/sty649
[arXiv:1612.06282 [astro-ph.CO]].


\bibitem{opendeSitter}
Y.~Yoon, ``Applying Verlinde's emergent gravity to the general FLRW universe, the open de Sitter universe,'' in preparation.






\end{thebibliography}
\end{document}